\documentstyle[11pt,newpasp,twoside]{article}
\def\edcomment#1{\iffalse\marginpar{\raggedright\sl#1\/}\else\relax\fi}
\reversemarginpar
\begin{document}
\title{High resolution simulations of star formation} 
\author{Henri M. J. Boffin} 
\affil{Royal Observatory of Belgium, 3 av. circulaire, 1180 Brussels}
\author{Neil Francis \& Anthony P. Whitworth} 
\affil{Dept. of Physics and Astronomy, Cardiff University, Wales}
\begin{abstract}
In order to study the capabilities of SPH in self-gravitating hydrodynamical problems, we have performed a series of three-dimensional numerical simulations of the collapse of a rotating homogeneous cloud, varying the number of particles and the artificial viscosity prescription.
\end{abstract}

The initial cloud is a sphere of radius $R=0.01635$ pc and mass $M=1$ M$_\odot$ in uniform rotation at 22.35 radMyr$^{-1}$ ($v=0.3566$ km/s). 
The cloud has uniform density, $3.739 10^{-18}$ gcm$^{-3}$ with a 10\%
$m=2$ perturbation. The temperature is 7.9 K, corresponding to a sound speed of 0.166 km/s. The ratios of the thermal and rotational energies to the magnitude of the gravitational potential energy are $\alpha = 0.26 $ and $\beta = 0.16$, respectively. We use a barotropic equation of state, which is isothermal up to a given density, where it becomes polytropic.
Our SPH code uses a tree to compute the gravity force and to find neighbours. The code is described in Boffin et al. (1998) and references therein. Individual time-steps are used.

It is already known (Bate \& Burkert 1997; Whitworth 1998) that, when 
studying self-gravitating flows with SPH, one needs always to resolve the Jeans mass, $M_J$. That is, if $N_n$ is the number of neighbours and $m_i$ the mass of a SPH particle $i$, one should always satisfy $N_n m_i < M_J$. 
In practice, Bate \& Burkert found that a Jeans mass should always be represented by at least twice this number.
As is common in most 3D SPH simulations, we use a value $N_n \sim 50$. We have performed simulations with $2~10^4, 6.3~10^4$ and $ 2.5~10^5$ particles.
In our 20,000 particle simulations, we thus resolve a Jeans mass of 
$\sim 0.005$ M$_\odot$, while in our 63,000 particles simulations, this
decreases to $\sim 1.6~10^{-3}$ M$_\odot$. In the $2.5~10^5$ particle simulation, we have taken $N_n \sim 200$, hence we resolve the same
Jeans mass as in the 63,000 particle simulations. We may however
expect that the increased number of particles has reduced the intrinsic
noise and  has also reduced  the artifical viscosity. Initially, the Jeans mass of the cloud is 0.2 M$_\odot$. In the 20,000 particle simulations, 
the simulation ceases to resolve the Jeans mass once the density
becomes greater than $6~10^{-16}$ gcm$^{-3}$. For the 63,000 and 250,000 simulations, this should in principle be 5.9 10$^{-14}$ gcm$^{-3}$, but because, from a density of 5 10$^{-14}$ gcm$^{-3}$ on, we use a barotropic equation of state where the sound speed scales as the 1.4 power of the density (hence the Jeans mass increases), we always resolve the Jeans mass in the  latter 
two simulations. 

In SPH, shocks are handled by artificial viscosity (A.V.). In our simulations, we have used the canonical values of $\alpha _{\rm SPH}=1$
and $\beta _{\rm SPH}=2$. However, it is well known that this A.V. also introduces shear viscosity. In order to try to make the A.V.  take effect only where it is needed, i.e. at the shocks, Morris \& Monaghan 
(1997) developed a viscosity switch, in which the value of $\alpha_{\rm SPH}$ is different for each particle and is time dependent.
In this scheme, $\alpha _{\rm SPH}$ is obtained from solving the equation
\begin{equation}
\frac{d\alpha_{\rm SPH}}{dt} = -\frac{\alpha_{\rm SPH} -\alpha_0}{\tau} + S,
\end{equation}
where $\alpha_0$ is a small arbitrary value, typically 0.1, to which 
$\alpha_{\rm SPH}$ converges after a decay time $\tau$, and $S$ is
the source term, which is here taken as the divergence of the velocity.
This prescription ensures that $\alpha_{\rm SPH}$ increases mostly at
the location of the shocks, which is what we require. We then take $\beta_{\rm SPH} = \alpha_{\rm SPH}$. In practice however, in the case of a collapsing flow, $\alpha_{\rm SPH}$ can reach large values even where there is no shock. We have therefore slightly modified this prescription, and {-- in the {\it modified switch} --} take as the source term, the divergence of the velocity only if the relative velocity is larger than the local sound speed. 
Our results may be summarized as follows:
\begin{itemize}
\item For the simulations with 20,000 particles, one can see spurious fragmentation inside the bar connecting the two cores of the proto-binary. With the viscosity switch, which in this case becomes too large and produces too much shear, the cores coalesce. This is avoided with the use of the modified switch.
\item The same behaviour is seen in the 64,000 particles simulations, in particular non-axisymmetric features with the standard A.V., and coalescence with the viscosity switch.
With the modified switch, the two cores separate from each others, and other fragments start to form in the accretion tails
\item In the 250,000 particles simulation, which was only run with the modified switch, the flow stays axisymmetric up to the end of our computations. There is a clear decrease in the noise. However, 
the increase in computer time required does not allow us to follow 
the simulation to such late stages as in the 63,000 particles case. Clearly, more clever schemes are required, such as particles splitting (Kitsionas et al., in preparation; see Whitworth, this volume).
\end{itemize}

\end{document}